\documentclass[epj]{svjour}
\usepackage{graphicx}
\usepackage{inputenc}
\usepackage{tablefootnote}
\usepackage{cite}
\usepackage{graphicx}
\usepackage{dcolumn}
\usepackage{bm}
\usepackage{braket}
\usepackage{hyperref}
\usepackage{amsmath}
\usepackage{fancyhdr}
\usepackage{float}
\usepackage{caption}
\usepackage{subcaption}
\usepackage{todonotes}
\usepackage{siunitx}
\usepackage{xcolor}
\usepackage{xspace}
\usepackage{ulem}

\newcommand{\Eotvos}{E\"otv\"os\xspace}

\newcommand{\keffi}[1]{k_{\text{eff,#1}}\xspace}

\newcommand{\mf}{m_F\xspace}

\newcommand{\stat}{\num{9E-8}\xspace}
\newcommand{\sys}{\num{3.1E-7}\xspace}

\newcommand{\Rbb}{\textsuperscript{85}Rb\xspace}
\newcommand{\Rb}{\textsuperscript{87}Rb\xspace}
\newcommand{\K}{\textsuperscript{39}K\xspace}

\newcommand{\Eot}{$\eta_{\,\text{Rb,K}}$\xspace}
\newcommand{\EotAB}{$\eta_{\,\text{A,B}}$\xspace}

\begin{document}
\title{Quantum test of the Universality of Free Fall using rubidium and potassium}
\titlerunning{Quantum test of the Universality of Free Fall using rubidium and potassium}
\author{H.~Albers\inst{1}, A.~Herbst\inst{1}, L.~L.~Richardson\inst{1}\thanks{\emph{Present address:} College of Optical Sciences, University of Arizona, Tucson, AZ 85721, USA}, H.~Heine\inst{1}, D.~Nath\inst{1}, J.~Hartwig\inst{1}, C.~Schubert\inst{1}, C.~Vogt\inst{2}, M.~Woltmann\inst{2}, C.~L\"{a}mmerzahl\inst{2}, S.~Herrmann\inst{2}, W.~Ertmer\inst{1}, E.~M.~Rasel\inst{1}, D.~Schlippert\inst{1}\thanks{Email: schlippert@iqo.uni-hannover.de}
}                     
\authorrunning{H.~Albers \textit{et al.}}
%
%
\institute{Leibniz Universit\"at Hannover, Institut f\"ur Quantenoptik, Welfengarten 1, 30167 Hannover, Germany \and ZARM Zentrum f\"ur angewandte Raumfahrttechnologie und Mikrogravitation, Universit\"at Bremen, Am Fallturm 2, \\28359 Bremen, Germany}
\date{Received: date / Revised version: date}
%
\abstract{
We report on an improved test of the Universality of Free Fall using a rubidium-potassium dual-species matter wave interferometer.
We describe our apparatus and detail challenges and solutions relevant when operating a potassium interferometer, as well as systematic effects affecting our measurement.
Our determination of the \Eotvos ratio yields \Eot~$=-1.9\times10^{-7}$ with a combined standard uncertainty of $\sigma_\eta=3.2\times10^{-7}$.
} 
\maketitle
\section{Introduction}
Matter wave interferometry is an effective toolbox to probe our understanding of nature.
Based on coherent manipulation of atomic ensembles, sensors capable of performing accurate inertial measurements have been demonstrated~\cite{Stockton11PRL,Menoret2018SciRep,Hu2013PRA,Savoie2018ScAdv,FreierJoP2016,Berg2015PRL,Sorrentino2012APL,Bidel2018NatComm,Geiger2011NatComm,Bidel2020JoG}.
These new atomic sensors allow accessing novel methods to understand fundamental physics~\cite{Rosi2014Nature,Jaffe2017NaturePhys,Haslinger2017NaturePhys,Bouchendira2011PRL,Parker2018Science}.

The Einstein equivalence principle (EEP) is a cornerstone for the theory of general relativity~\cite{Will06LivRev}.
It is composed of three components: Local Lorentz Invariance, Local Position Invariance, and the Universality of Free Fall.
A violation of any of the components would imply a violation of the EEP and could therefore yield modifications of general relativity with the possibility to reconcile it with quantum field theory and therefore form of a theory of quantum gravity.

The Universality of Free Fall (UFF) states the equality of inertial and gravitational mass $m_{\text{in}}=m_{\text{gr}}$ and implies that all objects freely falling in the same gravitational field experience the same acceleration. 
As a figure of merit for UFF tests in the Newtonian framework we can express differential acceleration measurements in the so-called \Eotvos ratio
\begin{equation}
\eta_{\,\text{A,B}}\equiv 2\thickspace\frac{g_{\text{A}}-g_{\text{B}}}{g_{\text{A}}+g_{\text{B}}}
=2\thickspace\frac{\left(\frac{m_{\text{gr}}}{m_{\text{in}}}\right)_{\text{A}}
-\left(\frac{m_{\text{gr}}}{m_{\text{in}}}\right)_{\text{B}}}
{\left(\frac{m_{\text{gr}}}{m_{\text{in}}}\right)_{\text{A}}
+\left(\frac{m_{\text{gr}}}{m_{\text{in}}}\right)_{\text{B}}}\;,
\label{eqn:eotvosratio}
\end{equation}
where A and B are the test masses, and $g_{\text{A,B}}$ is their respective local gravitational acceleration.

Tests of the UFF can be grouped in three categories depending on the nature of the test masses: i) classical, ii) semi-classical, and iii) quantum tests as reported in Table~\ref{tab:comparison}.
The UFF has been tested extensively by classical means, yielding the best uncertainty at parts in $10^{14}$.
In addition, since the first observation of a gravitationally induced phase in a matter-wave interferometer~\cite{Colella1975PRL}, a variety of quantum tests based on atom interferometry have emerged.
Due to their well-defined characteristics, isotopic purity, and by granting access to a novel range of species, they promise high sensitivity to possible violations of the EEP, e.g. when parametrizing observable physics in the minimal Standard Model extension~\cite{Kostelecky2011PRD,Hohensee2013PRL,Mueller2013arxiv} or in dilaton coupling scenarios~\cite{Damour2012CQG}.

In this article we report on an improved dual-species test of the Universality of Free Fall using laser-cooled \Rb and \K~\cite{Schlippert2014PRL}.
Improvements are mainly achieved by a better input state preparation for potassium yielding in an increased signal-to-noise ratio and longer integration time.
After a description of the experimental apparatus in Sec.~\ref{sec:apparatus}, we discuss our measurement scheme (Sec.~\ref{sec:measurements}) and close with a discussion of systematic effects (Sec.~\ref{sec:analysis}) affecting the measurement.
In Sec.~\ref{sec:conclusion} we present possible mitigation strategies and paths towards improved quantum tests of the UFF on ground~\cite{Dimopoulos07PRL,Overstreet18PRL,Zhou11GRG,Hartwig2015NJP} and in space~\cite{Aguilera2014CQG,Williams16NJP,Wolf2019arXiv}.

\begin{table*}[t]
\centering
\caption{Overview of UFF tests. 
We refer to experiments comparing the free fall acceleration of two isotopes of the same (different) chemical species as ``dual-isotope'' (``dual-species'').
Experiments comparing the free fall of different internal states of the same isotope are labelled ``dual-state''.
FCC -- Falling corner cube; AI -- Atom interferometer.}
\begin{tabular}{ l  c  c  r c}
\hline \hline
 
Experiment  & Test Masses & Type  & \Eotvos ratio \EotAB & Ref. \\ \hline

Capacitive accelerometers  & Ti -- Pt & Classical & $-0.1(1.3)\times10^{-14}$ & \cite{Touboul2017PRL} \\

Lunar laser ranging & Earth -- Moon & Classical &   $-3(5)\times10^{-14}$ & \cite{Hofmann2018CQG}\\

Torsion balance & \textsuperscript{9}Be -- Ti & Classical &   $0.3(1.8)\times10^{-13}$ & \cite{Schlamminger2008PRL} \\

Dual-FCC & Cu -- U & Classical  & $1.3(5.0)\times10^{-10}$ & \cite{Niebauer1987PRL}\\

\hline
\hline

FCC vs AI & SiO$_2$ -- \textsuperscript{133}Cs  & Semi-classical & $7(7)\times10^{-9}$ & \cite{Peters1999Nature} \\

FCC vs AI & SiO$_2$ -- \textsuperscript{87}Rb  & Semi-classical & $4.4(6.5)\times10^{-9}$ & \cite{Merlet2010Metrologia} \\

\hline
\hline

Dual-state AI  & \Rb & Quantum  & $0.9(2.7)\times10^{-10}$ & \cite{Zhang2018arxiv}\\

Dual-state AI  & \Rb & Quantum  & $1.4(2.8)\times10^{-9}$ & \cite{Rosi2017NatComm} \\

Dual-state AI  & \Rb & Quantum  & $0.2(1.2)\times10^{-7}$ & \cite{Duan2016PRL}\\

\hline

Dual-isotope AI & \Rbb -- \Rb & Quantum & $2.8(3.0)\times10^{-8}$ & \cite{Zhou2015PRL}\\

Dual-isotope AI & \Rbb -- \Rb & Quantum & $1.2(3.2)\times10^{-7}$ & \cite{Bonnin2013PRA} \\

Dual-isotope AI  & \textsuperscript{87}Sr -- \textsuperscript{88}Sr & Quantum & $0.2(1.6)\times10^{-7}$ & \cite{Tarallo2014PRL}\\
\hline

Dual-species AI  & \Rb -- \K & Quantum & $-0.3(5.4)\times10^{-7}$ & \cite{Schlippert2014PRL} \\

Dual-species AI  & \Rb -- \K & Quantum & $0.9(3.0)\times10^{-4}$ & \cite{Barrett2016NatComm} \\

\hline

Dual-species AI & \Rb -- \K & Quantum & $-1.9(3.2)\times10^{-7}$ & This work\\

\hline \hline

\end{tabular}
\label{tab:comparison}
\end{table*}

\section{Experimental apparatus}\label{sec:apparatus}
The experimental apparatus includes a vacuum system in which the atoms are interrogated, a laser system generating the light fields for manipulating the atoms, and optics for beam shaping and collecting fluorescence for detection at the vacuum chamber.
Below, these elements are described in more detail.
\subsection{Vacuum system}
\begin{figure}[t]
	\centering
    \includegraphics[width=.9\columnwidth]{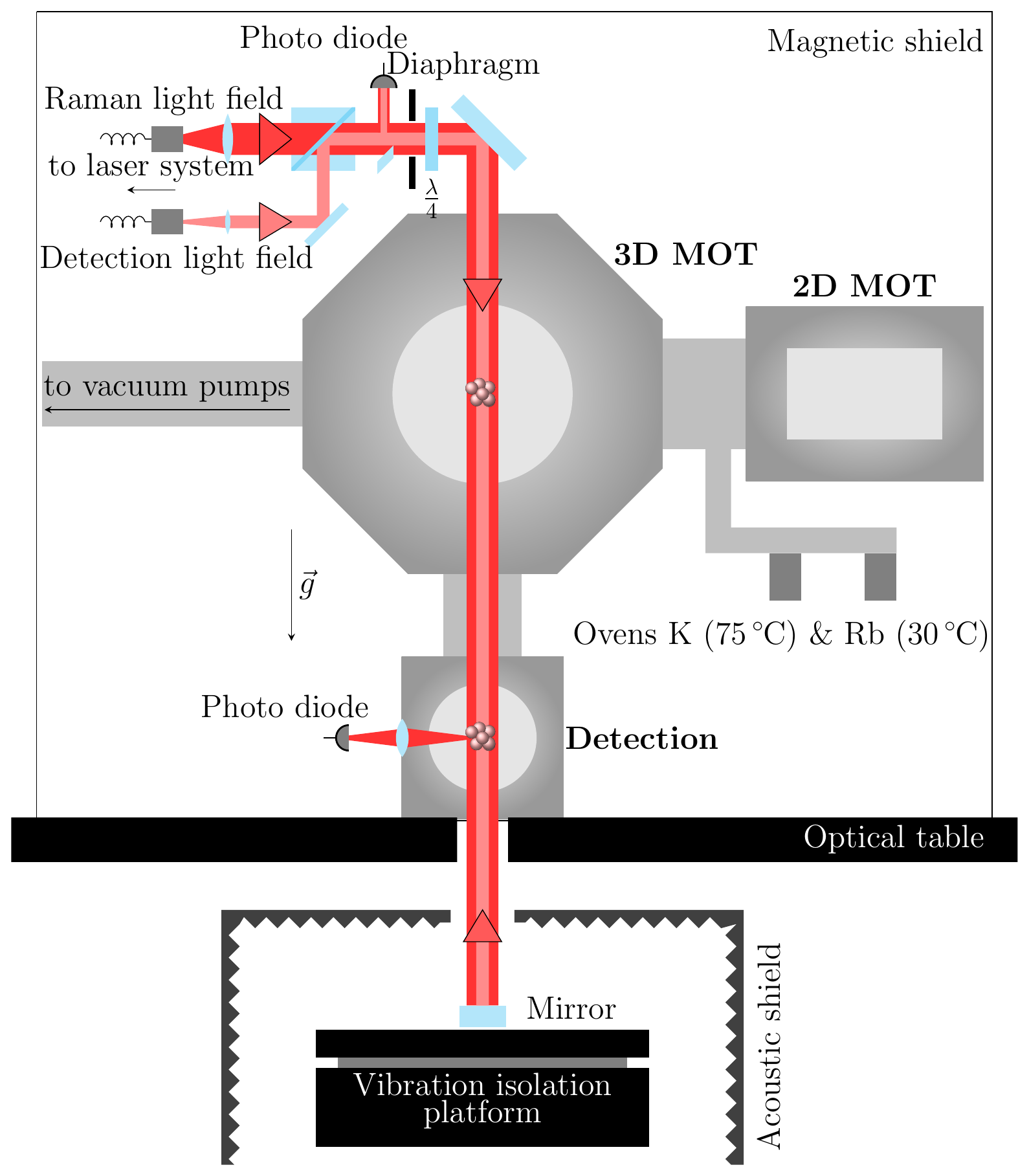}
	\caption{Schematic of the vacuum system and peripherals comprising the sensor head of the experimental setup. 
	The collimated light for the Raman pulses (dark red) and for detection (light red) are superimposed on a polarizing beam splitter which also cleans the Raman light polarization and allows for correcting detection intensity noise.
	Downstream, the light is shaped by a diaphragm to suppress reflections and unwanted diffraction from the viewport edge when passing the chamber and is circularly polarized afterwards.
	A retro-reflection mirror is situated on a vibration isolation platform in an acoustic isolation housing.
	Fluorescence readout is performed while the atoms fall through the large numerical aperture detection zone.}
	\label{img:setup}
\end{figure}
The vacuum system consists of three main parts as depicted in Fig.~\ref{img:setup} and is enclosed in a single-layer permalloy magnetic shield~\cite{Zaiser2010,Hartwig2013PHD}.
Cooling and trapping of \Rb and \K takes place in a double magneto-optical trap (MOT) setup comprising two custom-made aluminum chambers with indium-sealed viewports separated by a differential pumping tube.
Atoms are loaded into the 2D MOT from background vapor generated by ovens heated to \SI{30}{\degreeCelsius} (\SI{75}{\degreeCelsius}), yielding a partial pressure of \SI{1e-7}{\milli \bar}\\ (\SI{6e-6}{\milli \bar}) for rubidium (potassium).
A tube connects the 3D MOT chamber with a high aperture detection zone, allowing for \SI{200}{\ms} of free fall (\SI{19}{\cm} center to center).
\subsection{Laser system}
\begin{figure*}[t]
    \centering
	\includegraphics[width=0.9\textwidth]{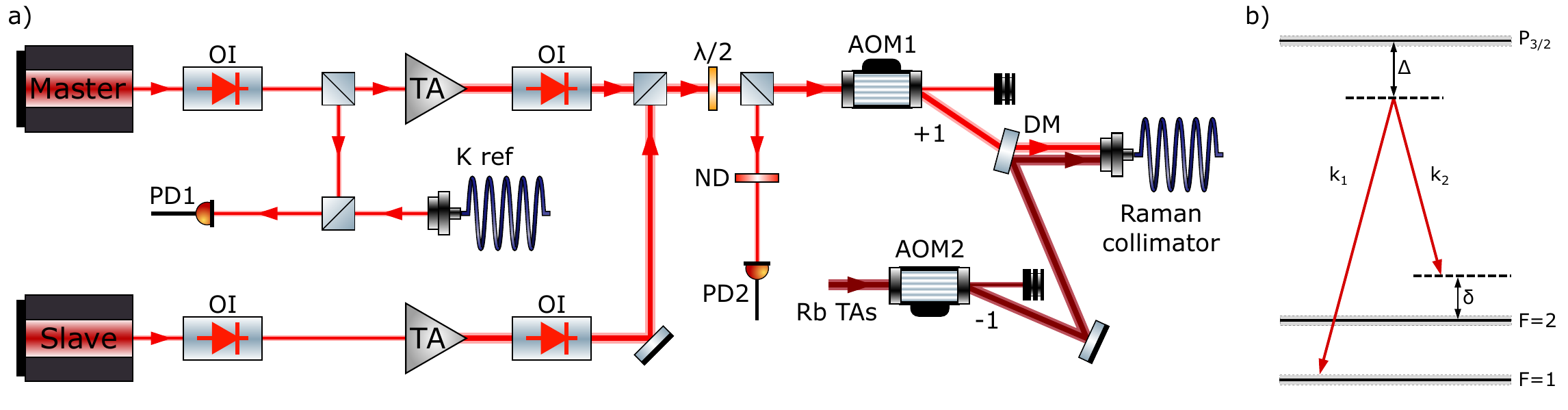}
	\caption{a) Schematic of the potassium Raman laser system (light red) and the superposition with the rubidium Raman laser light (dark red). Two ECDLs in a master-slave configuration are used for each species. 
	The master laser is offset locked (PD1) to the reference laser, while the slave laser is locked against the master (PD2). 
	Both beams are amplified using tapered amplifiers and can be switched with an AOM.  
	b) Level scheme of the Raman transition. A global detuning $\Delta$ is used for both lasers with an additional, variable detuning $\delta$ for the slave laser. 
	Abbreviations: OI -- Optical isolator, TA -- Tapered amplifier, AOM -- Acousto-optical modulator, PD -- Photo diode, DM -- Dichroic mirror, ND -- Neutral density filter.}
	\label{img:lasersystem}
\end{figure*}
For trapping and cooling of both atomic species we use the same laser system as in our previous work \cite{Schlippert2014PRL} and described in detail in \cite{Schlippert2014PHD, Hartwig2013PHD}.

In the following we refer to the $\ket{F=2} \rightarrow \ket{F'=3}$ as the cooling and the $\ket{F=1} \rightarrow \ket{F'=2}$ as the repumping transition.
Frequency references for the system are generated by two external cavity diode lasers (ECDL)~\cite{Baillard2006OC,Gilowski2007OC} which are stabilized to the $D_2$ line of rubidium (potassium) at \SI{780}{\nm} (\SI{767}{\nm}) by means of frequency modulation spectroscopy. 

The light fields for cooling and repumping of rubidium are generated by two ECDLs in a master-slave configuration. 
The repumping laser is stabilized to the reference laser and phase locked to be on resonance to the repumping transition. 
The cooling laser is phase-locked with respect to the repumper with an offset of $\SI{-3.1}{\Gamma}$ from the cooling transition, where $\Gamma\approx 2\pi\,\SI{6}{MHz}$ is the natural linewidth of rubidium and potassium.

The potassium cooling system consists of three independent ECDLs.
For the 2D MOT, one ECDL is phase locked to the reference and detuned by \SI{-1.3}{\Gamma} from the cooling transition. 
Repumping light is generated by passing this light through a double-pass acousto-optical modulator (AOM) operated at half the hyperfine transition frequency ($f_{\text{HFS}}\approx\SI{461}{MHz}$).
The radio frequency power is set to generate a 50:50 intensity ratio for cooling and repumping light.
The repumping light generated with this setup has a detuning of \SI{-4}{\Gamma} from resonance.
For the 3D MOT, two independent cooling and repumping lasers are phase locked to the reference laser with a variable detuning to provide the flexibility needed for the potassium sub-Doppler cooling scheme~\cite{Landini2011PRA}. 

All generated light fields except the rubidium repumper are amplified using tapered amplifiers (TA), while the intensity is controlled with AOMs.
Our setup yields cooling (C) and repumping (RP) intensities at the position of the atoms of $ I_{\text{C}} \approx 8\,I_{\text{sat}}, \,I_{\text{RP}} \approx 0.1\, I_{\text{sat}}$ ($I_{\text{C}} = I_{\text{RP}} \approx 12\,I_{\text{sat}}$) for rubidium (potassium), where $I_{\text{sat}}$ is each species's saturation intensity~\cite{Steck,Tiecke}.

To generate Raman beam splitter light, we utilize two additional ECDLs in master-oscillator power amplification (MOPA) configuration operated as a master-slave pair for each species. 
A schematic of the utilized system for potassium and the layout for superimposing the light with the rubidium Raman system is depicted in Fig.~\ref{img:lasersystem} a).
For the rubidium system a similar setup is used.
The master lasers are phase locked (PD1) to the reference lasers on the $\ket{F=1} \rightarrow \ket{F'=2}$ transition with a global detuning $\Delta$ of \SI{3.3}{\giga \Hz} for potassium, and \SI{1.6}{\giga \Hz} for rubidium (cf.~Fig.~\ref{img:lasersystem} b)).
To compensate for the Doppler shift, the slave lasers are phase locked (PD2) with a dynamic detuning $\delta$ to the master lasers on the $\ket{F=2} \rightarrow \ket{F'=2}$ transition.
The beam splitting light fields for both species can be switched independently using AOMs (AOM1 and AOM2 in Fig.~\ref{img:lasersystem}).
A dichroic mirror (DM) is used to superimpose the beam splitting light for both species.
Due to the small difference in wavelengths of the D$_2$ lines of \Rb and \K, we use common broadband optics at the experiment apparatus. 
Therefore we are able to generate a spatially and temporally overlapped cold atom cloud as well as superimposed Raman beams. 
\subsection{Interferometry and detection optics}
\begin{figure}[t]
	\centering
	\includegraphics[width=\columnwidth]{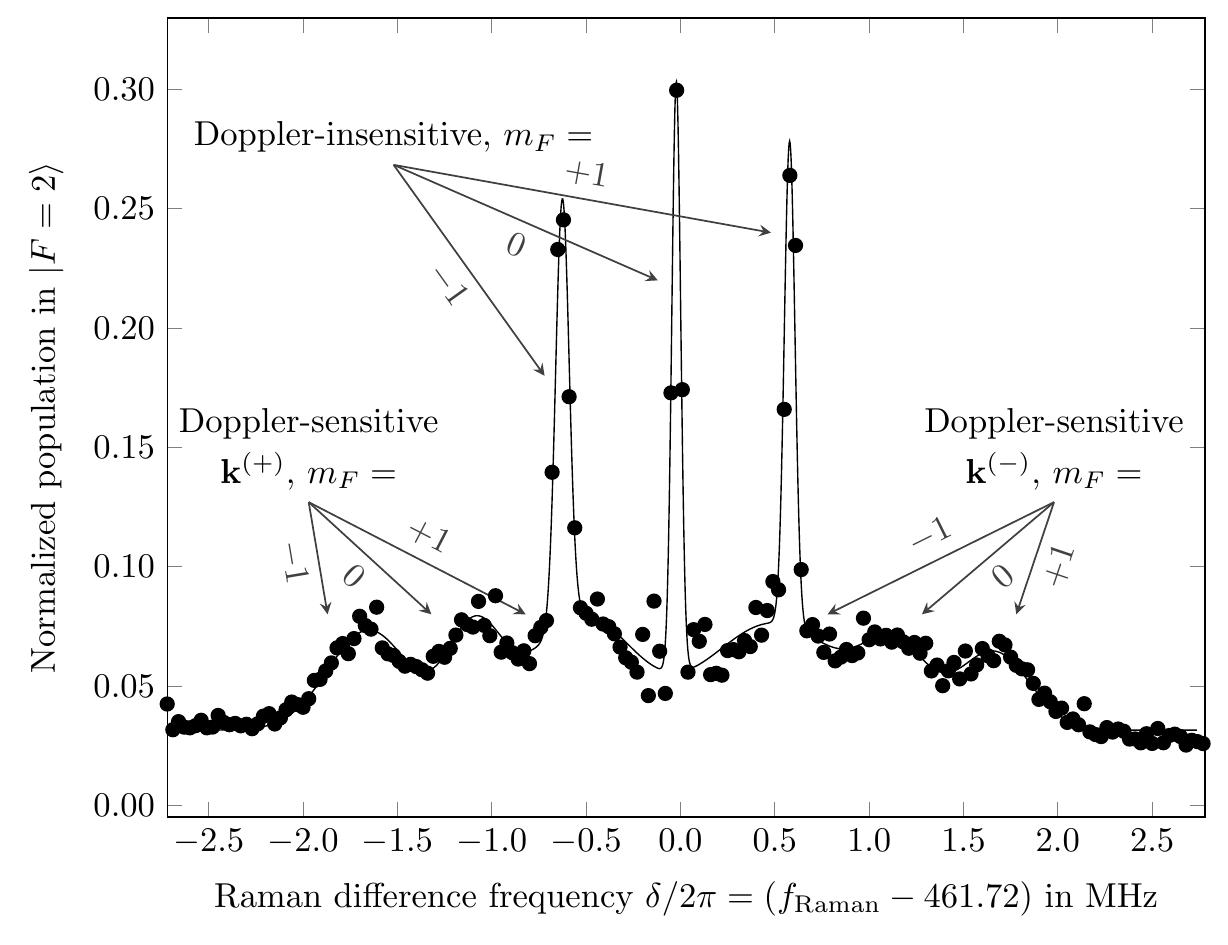}
	\caption{Typical potassium Raman resonance spectrum obtained by scanning $\delta$ and applying single Raman pulses. 
	The spectrum is acquired at the following parameters: pulse width $\tau=\SI{15}{\micro s}$, offset field $B_0=\SI{430}{mG}$, time of flight $t_{\text{TOF}}=\SI{43.25}{ms}$, temperature $T=\SI{32.6}{\micro K}$. 
	The solid black line is a guide to the eye.
	In the retro-reflected setup using $\sigma/\sigma$-polarization, a total of nine resonances are visible, three of which form one subset of Doppler-insensitive transitions. 
	The remaining two subsets of three Doppler-sensitive transitions each are labeled $k^{(+)}$ for upward and $k^{(-)}$ for downward momentum transfer. 
	Figure modified from Ref.~\cite{Schlippert2014PHD}.
	}
	\label{fig:ramanfreqscan_K}
\end{figure}
The Raman beams are set up in a retro-reflected  $\sigma^{+}-\sigma^{+}$ polarization configuration. 
They are collimated to a $1/e^2$-radius of 
$\sim$\SI{1.2}{\cm} using an achromatic lens ($f=\SI{100}{mm}$) and pass a cleanup polarization beam splitter, where they are superimposed with the detection light~(Fig.~\ref{img:setup}).
We obtain powers in the Raman master (slave) beam of  \SI{110}{\milli\watt} (\SI{110}{\milli\watt}) for potassium and \SI{45}{\milli\watt} (\SI{90}{\milli\watt}) for rubidium.
Both the detection and the Raman beams pass a diaphragm limiting the beam diameter such that no unwanted diffraction appears at the viewports.
A $\lambda/4$ retardation plate generates the circular polarization.
The beams are aligned parallel to gravity with two silver-coated mirrors.
Below the chamber the beams are retro-reflected by a mirror [Optique Fichou] with a $\lambda/20$ peak-to-valley flatness.
This mirror serves as the reference plane of the inertial measurements.
To suppress seismic noise, it is mounted on top of a benchtop vibration isolation platform [Minus-K BM-1], and the entire assembly is housed within a foam insulated acoustic isolation box.
For state-selective fluorescence detection, we utilize an optical system collecting fluorescence light with a large aperture lens ($f=\SI{50}{mm}$) and imaging it onto a photo diode [OSI Optoelectronics PIN-10D] in a $2f-2f$ configuration. 
\section{Measurements}\label{sec:measurements}
\subsection{Input state preparation}\label{sec:prep}
Initially, the atoms are loaded within \SI{1.3}{s} into the 3D-MOT. 
Subsequently, the magnetic fields are switched off for the optical molasses to enable sub-Doppler cooling.
Following the gray molasses method outlined in Ref.~\cite{Landini2011PRA} for \K and standard sub-Doppler cooling techniques for \Rb we typically obtain \SI{5e8} (\SI{6e7}) atoms at a temperature of \SI{21}{\micro \K} (\SI{28}{\micro \K}) for rubidium (potassium) within \SI{15}{\milli\second}.
Due to the trade-off in molasses temperature in favor of \K, the temperature of \Rb is higher than the typical value of $\SI{8}{\micro \K}$ when optimizing for \Rb only.

The procedure described in the following combines a state preparation with a vertical velocity selection for an increased signal-to-noise ratio~\cite{McGuirk2002PRA,Antoni-Micollier2017PRA}.
Subsequent to the sub-Doppler cooling the atoms are optically pumped into the $\ket{F=1,m_F}$ manifold.
Afterwards they are released into free fall.
A quantization field of $B_0 = \SI{500}{mG}$ is applied to lift the degeneracy of the magnetic sub-levels as depicted in a Raman resonance spectrum in Fig.~\ref{fig:ramanfreqscan_K}.
A microwave pulse transfers the atoms from the $\ket{F=1,m_F=0}$ into the $\ket{F=2,m_F=0}$ state. 
The microwave transitions are realized using a Yagi-Uda type antenna for potassium and a loop antenna for rubidium.
Then, the $\ket{F=1}$ state is depopulated by optically pumping the remaining atoms into the $\ket{F=2}$ manifold with an equal distribution.
This results in a population of the $\ket{F=2,m_F=0}$ state with up to 45 \% of the atoms~\cite{Antoni-Micollier2017PRA}.
After a time of flight of \SI{44}{ms} accommodating these steps, a velocity-selective Raman pulse selects a narrow vertical velocity class~\cite{Kasevich1991PRLb} of atoms from the $\ket{F=2,m_F=0}$ state by transferring them into the $\ket{F=1,m_F=0}$ state.
The remaining atoms from the $\ket{F=2}$ manifold are removed by a light pulse addressing the cooling transition, concluding the preparation sequence.
\subsection{Mach-Zehnder atom interferometry}
\begin{figure}[t]
\centering
\includegraphics[width=0.95\columnwidth]{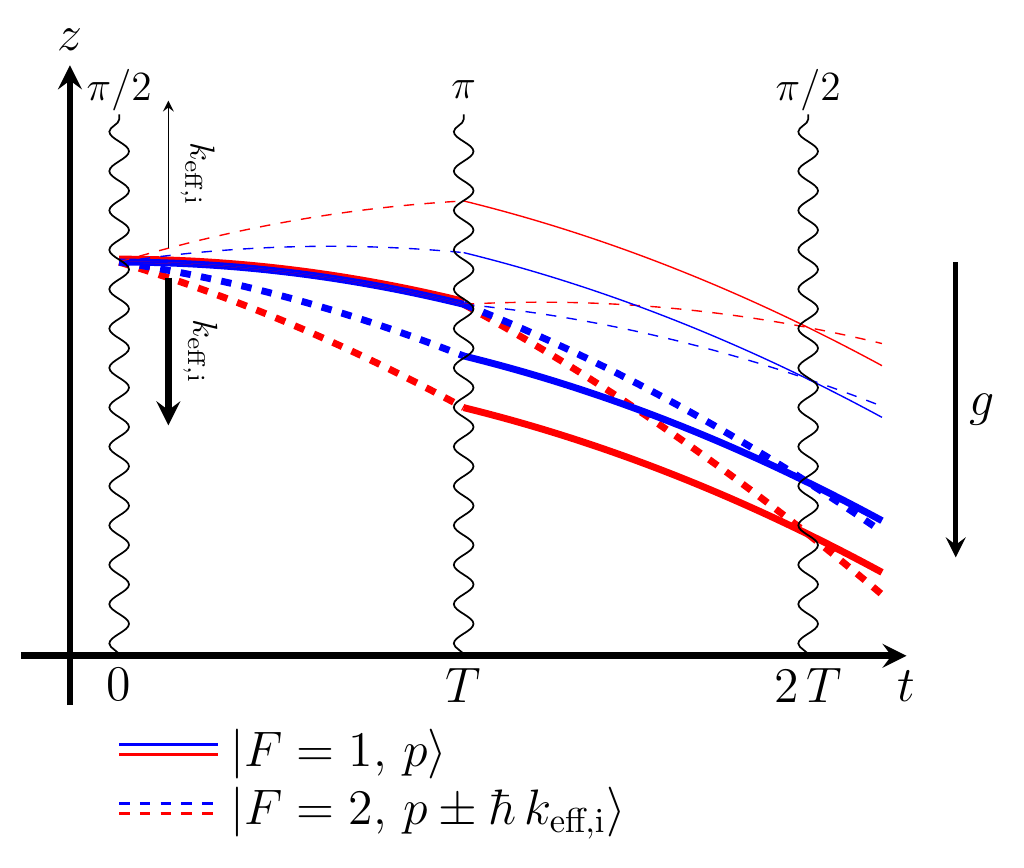}
\caption{Space-time diagram of a dual-species Mach-Zehnder matter-wave interferometer in a constant gravitational field for the downward (thick lines) and upward (thin lines) direction of momentum transfer.
Stimulated Raman transitions at times $0$, $T$, and $2\,T$ couple the states $\ket{F_i=1,\,p}$ (solid lines) and $\ket{F_i=2,\,p\pm\hbar\,\keffi{i}}$ (dashed lines), where $i$ stands for Rb (blue lines) or K (red lines). 
The velocity change induced by the Raman pulses is not to scale with respect to the gravitational acceleration.
}
\label{MZ}
\end{figure}

Atom interferometry is performed simultaneously with both species using the atomic sources described in Sec.~\ref{sec:prep}.
We employ two-photon Raman transitions~\cite{Kasevich1991PRL} driven by counter-prop\-agating beams with wave vectors $k_i=2\pi/\lambda_i$, where $\lambda_i$ refers to the D$_2$ transition wavelength.
The index $i$ indicates the species \Rb and \K.
We form a Mach-Zehnder-type atom interferometer with a $\pi/2-\pi-\pi/2$ pulse sequence separated by free evolution times $T$ to coherently split, reflect, and recombine the wavepackets.
The atomic recoil $\Delta p \approx 2 \hbar k_i = \hbar k_{\text{eff},i}$ induced by atom-light interaction leads to a finite space-time area enclosed by the AI~(Fig. $\ref{MZ}$).
For our scale factors, the presence of our commercial vibration isolation platform allows us to scan fringes as opposed to using an ellipse fitting algorithm commonly used in differential atom interferometers~\cite{Varoquaux_2009,Chen_2014,Barrett2015NJP}.

The population of the output ports of the interferometer depends on the accumulated phase difference $\Delta \phi$ between the two paths of the interferometer~\cite{Kasevich1992, Peters1999Nature, Berman1997} and is given by:
\begin{equation}
P_{\ket{F=2}} = A \cdot \cos(\Delta \phi) + P_0\; ,
\label{eqn:population2phase}
\end{equation}
where $P_{\ket{F=2}}$ is the fraction of atoms in the excited $\ket{F=2}$ state, $C=A/P_0$ is the contrast, and $P_0$ the offset.
The population is measured by a normalized state selective fluorescence detection, within which the pulses reading out potassium are nested within the rubidium detection sequence.\\
The leading order phase shift of such an interferometer due to an acceleration\footnote{If derived from the Schr\"odinger equation with masses $m_{\text{in}}$ in the kinetic term and $m_{\text{gr}}$ in the Newtonian potential, a prefactor resembling those in Eq. \ref{eqn:eotvosratio}, $a_i\rightarrow \frac{m_{\text{gr}}}{m_{\text{in}}}\,a_i$ becomes apparent.} $a_i$ in the direction of beam splitting reads~\cite{Kasevich1992, Peters1999Nature, Berman1997}
\begin{equation}
\Delta \phi = k_{\text{eff},i}\,a_i\,T^2 \; .
\label{deltaphi}
\end{equation}
Applying a phase-continuous frequency ramp $\alpha_i$ not only maintains the Raman resonance condition under the influence of a gravitational Doppler shift, but also mimics an effective acceleration of the Raman wave fronts
\begin{equation}
a_i = \frac{\alpha_i}{k_{\text{eff},i}}
\label{accalphag}
\end{equation}
and accordingly enters the phase shift as follows:
\begin{equation}\label{eq:phaseshift}
    \Delta\phi_i=(g_i-\frac{\alpha_i}{\keffi{i}})\cdot\keffi{i}\cdot T^2\;.
\end{equation}
For $\alpha_i = k_{\text{eff},i} \cdot g_i$ the accumulated phase shift $\Delta \phi_i = 0$ for all pulse separation times $T$.

We apply the momentum reversal technique~\cite{McGuirk2002PRA, Louchet-Chauvet2011} to suppress systematic errors independent of the direction of momentum transfer.
We distinguish two types of undesired phase perturbations, $k$-dependent ($\delta\phi_{dep}$) and $k$-independent ($\delta\phi_{ind}$) shifts.
In our setup, the two possible counter-propagating Raman beam configurations have opposite effective wave vectors and allow for selecting the direction of momentum transfer. 
We label these particular transitions as $\mathbf{k^{(+)}}$ and $\mathbf{k^{(-)}}$~(Fig.~\ref{fig:ramanfreqscan_K}).
\begin{figure*}
    \centering
    \includegraphics[width=1\textwidth]{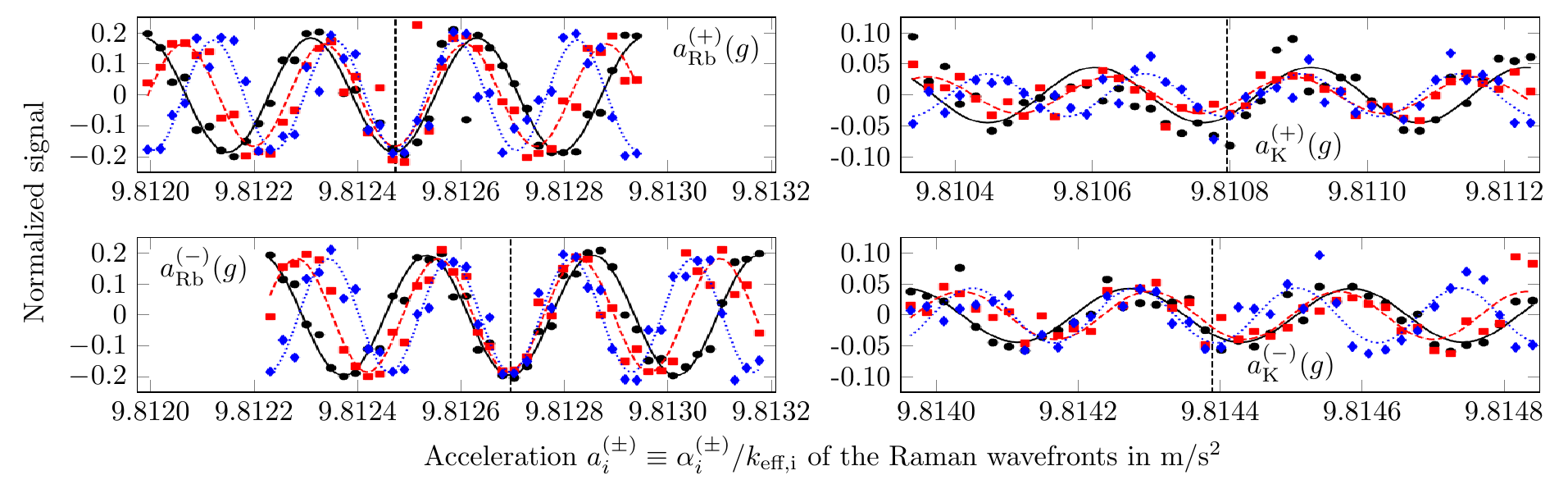}
    \caption{Determination of the differential gravitational acceleration of rubidium (left) and potassium (right). Typical fringe signals and sinusoidal fit functions are plotted in dependence of the effective Raman wave front acceleration for pulse separation times $T=\SI{35}{ms}$ (black circles and solid black line), $T=\SI{38}{ms}$ (red squares and dashed blue line), and $T=\SI{41}{ms}$ (blue diamonds and dotted red line) for upward $(+)$ and downward $(-)$ direction of momentum transfer. 
    The central fringe positions $a_i^{(\pm)}(g)$, $i=\text{Rb,K}$ for $T=\SI{41}{ms}$ are marked with dashed vertical lines. 
    The data sets are corrected for slow linear drifts caused by varying offsets in the detection and global signal offsets.}
    \label{fig:uff}
\end{figure*}
The phase shifts in $\mathbf{k^{(+)}}$ and $\mathbf{k^{(-)}}$ configuration can be written as:
\begin{align}
 \Delta\phi_{+} &= k_{\text{eff}} a T^2 + \delta\phi_{ind} + \delta\phi_{dep}\\
 \Delta\phi_{-} &= -k_{\text{eff}} a T^2 + \delta\phi_{ind} - \delta\phi_{dep}
 \label{eqn:kreversalphase} 
\end{align}
Consequently, their phase difference is given by:
\begin{equation}
 \Delta\phi_{tot}=\frac{\Delta\phi_{+}-\Delta\phi_{-}}{2} = k_{\text{eff}} a T^2 + \delta\phi_{dep}\; .
 \label{eqn:kreversalfinalphase}
\end{equation}
Hence, by alternating the direction of momentum transfer we can largely suppress momentum independent ($\delta\phi_{ind}$) systematic effects, e.g. the AC-Stark shift, with dynamics slower than a typical momentum reversal sequence as described in the following subsection.
\subsection{Obtaining the \Eotvos ratio}
The gravitational accelerations $g_i$~(Eq.~\ref{eq:phaseshift}) are determined through the central fringe positions $a_i^{(\pm)}(g)$.
For determining the latter, we operate both interferometers at three pulse separation times $T=35,\,38,\,41\,\text{ms}$.
Fig.~\ref{fig:uff} displays scans around the respective $a_i^{(\pm)}(g)$.
Here, for the downward direction of momentum transfer the sign of the phase shift is inverted in order to yield a positive value $g_i>0$.

We then operate both interferometers simultaneously around their central fringe positions with $T=\SI{41}{ms}$.
To this end we scan across the central fringe positions in 10 steps and alternate the direction of momentum transfer afterwards.
This procedure constitutes a single measurement cycle with a duration of \SI{32}{s} ($2\times10$ shots). 
Each measurement cycle yields $g_{\text{Rb}}$ and $g_{\text{K}}$, allowing us to compute an \Eotvos ratio ~(Eq.~\ref{eqn:eotvosratio}).
\section{Data analysis and results}\label{sec:analysis}
\subsection{Statistical analysis}
\begin{figure}
    \centering
    \includegraphics[width=\columnwidth]{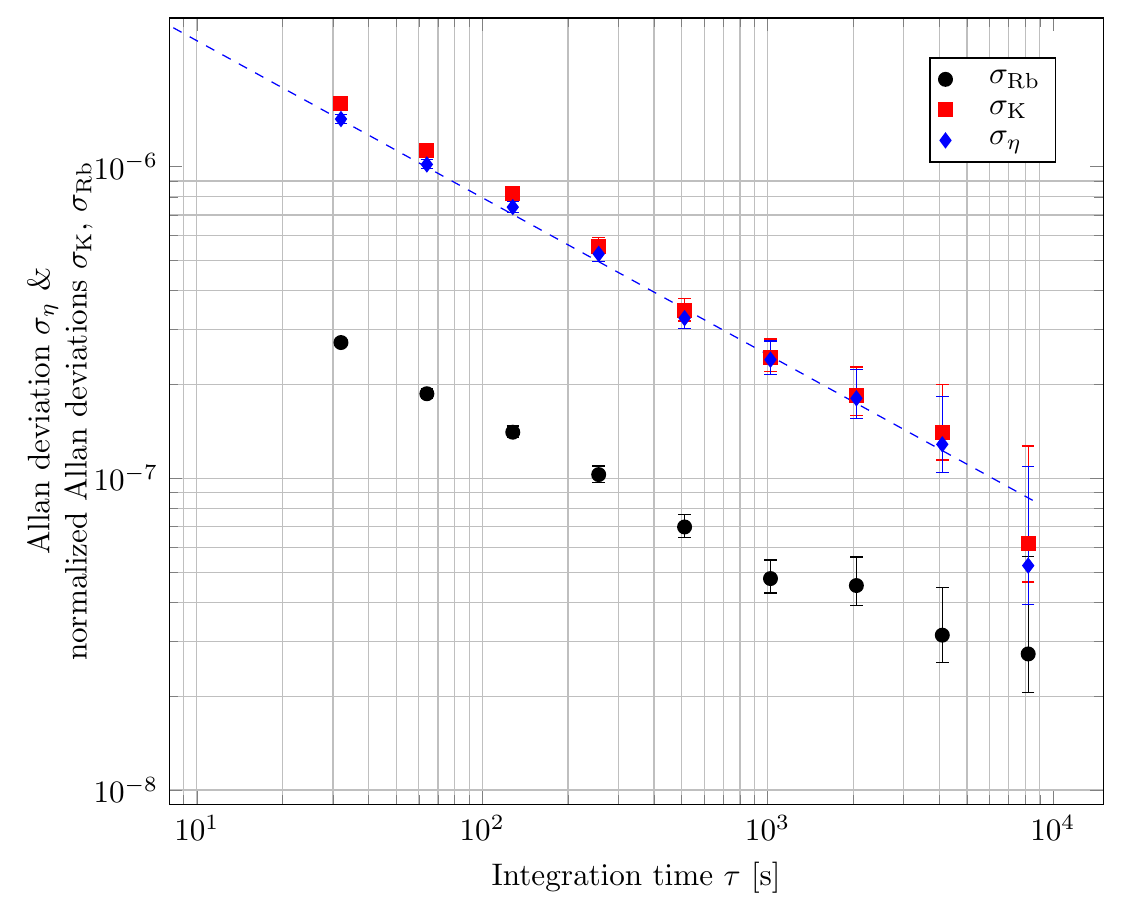}
    \caption{Normalized Allan deviations $\sigma_{\text{Rb}}$, $\sigma_{\text{K}}$, and the Allan deviation $\sigma_\eta$ of the signals providing us with the accelerations $g_{\text{Rb}}$ and $g_{\text{K}}$ of rubidium (black circles) potassium (red squares), and of the \Eotvos ratio \Eot~(blue diamonds) in their dependence of the integration time $\tau$. Using a fit $\propto 1/\sqrt{\tau}$ (blue dashed line) we extract a statistical uncertainty of the \Eotvos ratio of $\sigma_\eta=\stat$ after \SI{8192}{s} integration. The measurement is solely limited by the stability of the potassium signal.}
    \label{fig:adev}
\end{figure}
We acquire 30000 shots over consecutive 13 hours. 
This is limited by technical circumstances related to the stability of the laser locks.
Fig.~\ref{fig:adev} shows the normalized Allan deviation of our measurement which yields an instability $\sigma_\eta=\stat$ after 8192\,s integration time.
This instability is fully dominated by the potassium interferometer due to its larger technical noise influence stemming mainly from the detection at a significantly lower contrast~(Fig.~\ref{fig:uff}).
The latter can be explained by the underlying large transverse expansion rate of potassium and related homogeneous excitation when the ensemble diameter becomes comparable to the Raman beam diameter.
\subsection{Systematic effects}\label{systematics}
Table~\ref{tab:systematics} lists the systematic effects that are not suppressed by the momentum reversal method.
Here, the analysis follows Refs.~\cite{Dimopoulos2008PRD,Gauguet2008PRA,Schubert2013arxiv,Louchet-Chauvet2011,Peters2001Metrologia,Dubetsky2019} with the following assumptions for Rb (K): 
Temperature -- \SI{21}{\micro K} (\SI{28}{\micro K}) with a \SI{10}{\percent} uncertainty; 
Initial size -- \SI{1}{mm} (\SI{1}{mm});
$\pi$-pulse width -- \SI{15}{\micro s} (\SI{15}{\micro s});
Free evolution time -- \SI{41}{ms} (\SI{41}{ms}); 
Time-of-flight prior to the 1\textsuperscript{st} interferometry pulse -- \SI{54.5}{ms};
Differential center-of-mass (COM) position uncertainty in $z$ -- \SI{1}{mm};
Differential COM velocity uncertainty in $z$ -- \SI{1}{mm/s}.

Below we discuss the treatment of the dominant systematic contributions originating from stray magnetic fields and wave front aberration.
\begin{table}
\centering
\caption{Estimated bias contributions for the \Eot ratio and their uncertainties $\sigma$. We estimate the uncertainties to be uncorrelated at the discussed level of accuracy.}
\begin{tabular}{ c  c  c }
\hline \hline
 
 Contribution  & Correction $\Delta \eta $ & Uncertainty $\delta \eta$  \\ \hline

Zeeman effect & $-1.3   \times 10^{-6} $ & $6.0 \times10^{-8}$ \\

Wave front aberration  & 0 & 3 $ \times 10^{-7}$ \\

Coriolis force & 0 & 9 $ \times 10^{-9}$ \\

2-photon light shift & 3.08 $\times 10^{-8}$ & 6 $\times 10^{-10}$ \\

Effective wave vector & 0 & 1.3 $\times 10^{-9}$ \\

1\textsuperscript{st} order gravity gradient & 0 & 1 $\times 10^{-10}$ \\
\hline
Total bias &-1.28 $\times 10^{-6}$ & 3.1 $\times 10^{-7}$\\

\hline \hline

\end{tabular}
\label{tab:systematics}
\end{table}
\subsubsection{Zeeman effect}\label{sec:zeeman}
Magnetic fields along the interferometric trajectories change each respective species's hyperfine transition frequency. 
Due to state preparation into the $\ket{F=1,\mf=0}$ magnetically insensitive state the Zeeman effect cancels to first order.
However spatial and temporal variations of the magnetic field along the axis of interferometry contribute a non-zero bias phase resulting from the remaining clock shift affecting atoms in $\mf=0$.

Magnetic fields shift the hyperfine transition frequency by $\Delta\omega^{i}_{\text{clock}} =2 \pi \kappa^i \cdot B^2 $, where $\kappa^{Rb}=\SI{575.15}{Hz/G^2}$  for \Rb, and $\kappa^{K}=\SI{8.5}{kHz/G^2}$  for \K. 
Based on a characterization using Raman spectroscopy at different positions along the vertical axis we can model the magnetic field as
\begin{equation}
B(z(t),t) =  B_0(t) + \frac{\partial B }{\partial z} \cdot z_{\text{COM}}(t)
\end{equation}
where $B_0(t)$ describes the temporal behaviour of the bias field due to the switching behaviour of the respective power supply\footnote{We have characterized the switch-on to be saturating at \SI{500}{mG} with a time constant of \SI{90}{ms}.}, and $z_{\text{COM}}(t)$ is the free fall center of mass motion of the atoms defined as: $z_{\text{COM}}(t)= z_0 + (v_0 \pm v_{\text{rec}}/{2})\cdot t+1/2 at^2$ with $z_0$ being the initial position of the atoms, $v_0$ the velocity, and $v_{\text{rec}}$ the recoil velocity defined as $v_{\text{rec}}=\hbar k_{\text{eff}}/{m}$.
Only the recoil velocity is dependent on the direction of momentum transfer, and therefore all other components are suppressed by $k$--reversal. 
Using the sensitivity function formalism~\cite{Cheinet2008IEEE} we can calculate the frequency shifts
\begin{equation}
\Delta\omega_{\text{Z}^{(\pm)},\,i}\,(t)\equiv\pm\,2 \pi \kappa^i\cdot \frac{\partial B}{\partial z}\cdot B\,(t)\cdot v_{\text{rec},\,i}\,t\;,
\end{equation}
with the clock shift $\Delta\omega_{\text{clock},\,i}$. 
Computing the integral of the clock shift weighted with the sensitivity function $g_s(t)$
\begin{equation}\label{eq:recoilshift}
\Delta\Phi_i^{\text{Z}}\equiv\int\limits^{\infty}_{-\infty}g_{s}\,(t)\;\Delta\omega_{\text{Z}^{(\pm)},\,i}\,(t)\,\mathrm{d}t
\end{equation}
allows deriving the bias due to the Zeeman effect.
With a gradient $\frac{\partial B }{\partial z}=\SI{3}{mG/cm}$ and using the sensitivity formalism for rubidium and potassium the inferred bias affecting the \Eotvos ratio amounts to $-1.30(0.06)\times 10^{-6}$.
We note that the resulting bias was also confirmed using a perturbation theory formalism~\cite{Ufrecht2019}.
\subsubsection{Wave front aberration}
Our estimation of systematic uncertainty owing to wave front aberration is based on numerical simulation.
To this end, we take into account the Raman light fields' propagation including curvatures of view ports and the retro-reflection mirror, uncertainty in the positioning of the collimation lens, and differential ensemble expansion.
In $10^4$ trials, we randomly vary these parameters as follows and calculate the resulting phase contribution~\cite{Schubert2013arxiv}: 
For the top and bottom view port curvatures~($\lambda/10$) and the retro-reflection mirror~($\lambda/20$) we assume uncertainties of \SI{10}{\percent}, for the positioning of the collimation lens we assume an uncertainty of \SI{0.1}{\percent} and the ensemble temperatures are varied with an uncertainty of \SI{10}{\percent}.
Statistical analysis then yields an uncertainty in the \Eotvos ratio of \SI{3e-7}{} due to wave front aberration.
\subsection{Summary \& discussion}
We determine an \Eotvos ratio of \Eot~$=-1.9\times10^{-7}$ with a combined (statistical and systematic) standard uncertainty of $3.2\times10^{-7}$, constituting about a factor of two improvement over our previous result~\cite{Schlippert2014PRL}.
We estimate a contribution of the statistical uncertainty of \stat and the systematic uncertainty of \sys.

Increased free fall times in comparison to our previous result~\cite{Schlippert2014PRL} lead to larger systematic error contributions, e.g., due to the distance traveled through the magnetic field gradient which yields an increased correction, and more challenging characterization of it resulting in a larger uncertainty.
The increase in the contribution from wave front aberration is caused by a more defensive modeling of all relevant optical components.
\section{Conclusion \& outlook}\label{sec:conclusion}
We reported on a test of the UFF with our dual atom interferometer operating simultaneously with \Rb and \K.
The result of our measurement of the \Eotvos parameter is $-1.9(3.2)\times10^{-7}$ with the lowest uncertainty in an atom interferometer with two different chemical elements reported so far.
It corresponds to an improvement of a factor of two with respect to our previous result which we attribute to an improved state preparation of the \K ensemble.\\
Our analysis shows that the intrinsic noise of the \K interferometer limits the statistical uncertainty.
The reasons are the residual transverse expansion rate implying imperfections in the coherent excitation leading to a reduction in contrast.
Among the systematic effects, we identify the inhomogeneity in the magnetic bias field and wave front aberration as the main contributors.\\
Advanced cooling techniques such as evaporation in an optical dipole trap~\cite{Salomon2014PRA} are expected to enhance the contrast~\cite{Abend2016PRL,Szigeti12NJP} and will allow us to reduce wave front-related errors by achieving colder temperatures and by tuning the differential expansion of the two species.
The homogeneity of the magnetic field can be improved by an upgrade of the magnetic shield~\cite{Wodey2019arxiv}, a more in-depth characterization, and advanced center-of-mass control over the ensembles.
By relying on the differential suppression of vibration noise between the two elements~\cite{Barrett2015NJP} we envisage the perspective for a test on the level of $10^{-9}$.\\
To date, the universality of free fall has proven to be successful with no precision test uncovering a significant deviation.
Atom interferometers add a complementary approach to the toolbox for these tests.
Compared to the best classical tests~(Table~\ref{tab:comparison}), this still represents a modest result, but further enhancements are possible and realistic.
Using evaporated atoms and matter-wave collimation techniques~\cite{Rudolph16,Kovachy15PRL,Muntinga2013PRL} opens the pathway to upgrade the beam splitters to hundreds of coherent photon momentum kicks~\cite{Gebbe2019arXiv,Plotkin2018PRL,Jaffe2018PRL,Kotru2015PRL,McDonald2014EPL,Chiow2011PRL,Chiow2009PRL,Clade2009PRL} and extended free evolution times on the order of seconds.
Very long baseline atom interferometers relying on these techniques promise to venture beyond $10^{-13}$~\cite{Overstreet18PRL,Hartwig2015NJP,Zhou11GRG,Dimopoulos07PRL}.
In parallel, microgravity research investigates the benefits and the adaption of atom interferometers to operation in drop tower experiments~\cite{Kulas2017MGST,Vogt2020PRA,Muntinga2013PRL}, on a sounding rocket~\cite{Becker2018Nature}, the international space station~\cite{Williams16NJP,Tino2013NuclPhysB}, and dedicated satellite missions~\cite{Wolf2019arXiv,Aguilera2014CQG} with the target of $10^{-15}$ and beyond.
\section{Acknowledgments}
We are grateful to \'E. Wodey, H. Ahlers, and B. Barrett for discussions and critical proof reading of the manuscript, and T. Hensel for careful cross checks concerning the systematic error originating from magnetic fields.
This project is supported by the German Space Agency (DLR) with funds provided by the Federal Ministry for Economic Affairs and Energy (BMWi) due to an enactment of the German Bundestag under Grant No. DLR 50WM1641 (PRIMUS-III).
The authors furthermore acknowledge financial support by the German Science Foundation (DFG) through the Collaborative Research Centers 1128 ``geo-Q'' (projects A02 and F01), 1227 ``DQ-mat'' (project B07), the Deutsche Forsch\-ungs\-gemeinschaft (DFG, German Research Foundation) under Germany’s Excellence Strategy – EXC-2123 QuantumFrontiers – 390837967 (research unit B02), ``Nieders\"achs\-isches Vorab'' through the ``Quan\-tum- and Nano\-Metrology'' (QU\-ANO\-MET) initiative within the project QT3, and ``Wege in die Forschung (II)'' of Leibniz Universit\"at Hannover. 
A.H. and D.S. gratefully acknowledge funding by the Federal Ministry of Education and Research (BMBF) through the funding program Photonics Research Germany under contract number 13N14875.
\section{Authors contributions}
W.E.,
E.M.R.,
C.S.,
J.H.,
and D.S. designed the atom interferometer and its laser system.
L.L.R.,
H.A.,
D.N.,
J.H.,
and D.S. contributed to the design of the atom interferometer and its laser system and realised the overall setup.
H.A.,
L.L.R.,
H.H.,
and D.N. operated the final experimental setup.
H.A.,
A.H.,
D.S.,
J.H.,
C.S.,
and H.H. performed the analysis of the data presented in this manuscript.
D.S.,
H.A.,
L.L.R.,
C.S.,
and A.H. drafted the initial manuscript.
C.V.,
M.W.,
C.L.,
and S.H. provided major input to the manuscript 
and all authors critically reviewed and approved of the final version.
%
\bibliographystyle{unsrt}
\bibliography{main}

\end{document}